\documentclass[%
 reprint,
 amsmath,amssymb,
 aps,
]{revtex4-2}

\usepackage{graphicx}
\usepackage{dcolumn}
\usepackage{bm}
\usepackage{ragged2e}

\begin{document}

\preprint{APS/123-QED}

\title{Velocity space compression from Fermi acceleration with Lorentz scattering}

\author{J. C. Waybright}
\author{M. E. Mlodik}%
\author{N. J. Fisch}
\affiliation{%
 Department of Astrophysical Sciences, Princeton University, Princeton, NJ 08540
}%

\date{\today}

\begin{abstract}
The Fermi acceleration model was introduced to describe how cosmic ray particles are accelerated to great speeds by interacting with moving magnetic fields. We identify a new variation of the model where light ions interact with a moving wall while undergoing pitch angle scattering through Coulomb collisions due to the presence of a heavier ionic species. The collisions introduce a stochastic component which adds complexity to the particle acceleration profile and sets it apart from collisionless Fermi acceleration models. The unusual effect captured by this simplified variation of Fermi acceleration is the non-conservation of phase space, with the possibility for a distribution of particles initially monotonically decreasing in energy to exhibit an energy peak upon compression. A peaked energy distribution might have interesting applications, such as to optimize fusion reactivity or to characterize astrophysical phenomena that exhibit non-thermal features. 
\end{abstract}

\maketitle


\section{\label{sec:level1} Introduction}

The Fermi acceleration model was introduced to describe how cosmic ray particles are accelerated to great speeds by interacting with moving magnetic fields \cite{Fermi49}. Since then, many variations of the model have been studied. One well known example is the Fermi-Ulam model which describes the acceleration of an ensemble of non-interacting particles bouncing between a moving wall and a stationary wall \cite{Ulam61}. Several studies have examined different billiard shapes \cite{Gelfriech12, Zhou2020, Lenz08, Leonel09}, wall movement setups \cite{Gelfriech12, Jarzynski93, Loskutov2000, Gelfreich2008}, and particle forces \cite{Leonel_2005, Silveira21} and how this affects particle acceleration and accessible points in phase space.  \\
\indent Consider another variation where particles interact with a moving wall while also undergoing pitch angle scattering. Suppose that the pitch angle scattering, also called Lorentz scattering, is effected by means of Coulomb collisions of light ions with a background of heavy ions. We assume that other types of collisions occur on a much greater timescale, and therefore do not consider their effects in our model. We also assume that the moving wall has a negligible effect on the density of the heavy ions either by allowing these particles to pass through, or stick to the wall as it compresses. Fig. 1 shows a graphic of this system. This setup is not unique in that it considers Fermi acceleration with pitch angle scattering, as other studies have investigated aspects of this collisional effect \cite{Chandran00, Selkowitz04, Liu17, Scott78}. However, these studies have added many features simultaneously such as electromagnetic fields, fluid effects, and complex scattering systems which do not isolate the effects we report in this paper. \\
\indent The unique aspect of our problem is the simplicity of our system, paired with the limits in which we study the collisions, which results in a distinctive scaling between the change in energy and the initial energy of a particle. Since we are setting the Coulomb collision time as the smallest time scale in the system, pitch angle scattering will be the primary mechanism for reflecting particles back towards the moving wall, rather than being reflected by any wall on the other side.  Introducing this stochastic effect into the system will influence the frequency at which particles interact with the moving wall, and therefore also affect the total evolution of the particle distribution function. In particular, due to the relationship between the mean free path for Coulomb collisions and the speed of a particle, the rate at which a particle is accelerated by the moving wall may be heavily dependent on its initial speed and the collision frequency with the background species. This would imply that such a system could be tuned with these parameters to accelerate distributions of particles in a desired way to achieve a peaked energy distribution.
\begin{center}
    \includegraphics[scale=0.65]{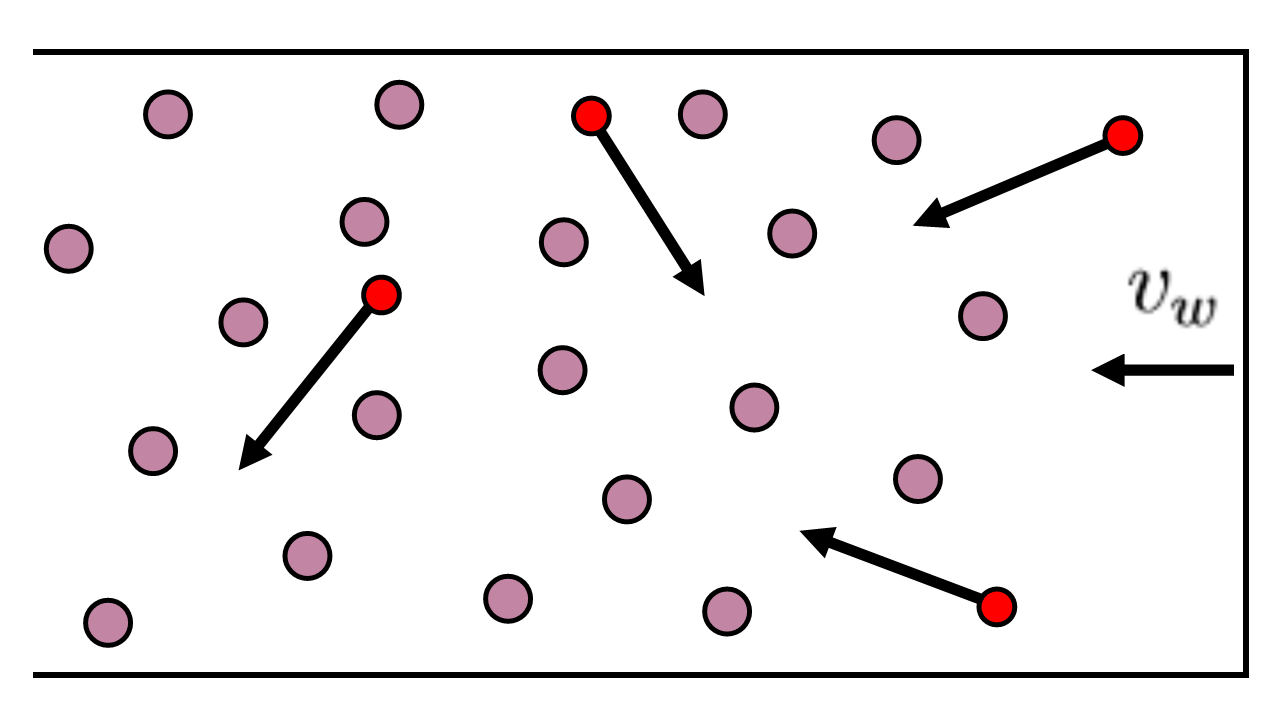} \\
    \begin{justify}
    Figure 1: Particles interacting with a wall moving at speed $v_w$ while pitch angle scattering with a background species.
    \end{justify}
\end{center}
\indent \indent To get an idea of the underlying physics in the system, first we study a simpler 1-D problem where we treat Lorentz scattering by having particles reflected back towards the moving wall after traveling one collisional mean free path. We call this the plasma wall approximation, since particles are being reflected at a fixed length. This setup results in a new invariant conserved during compression which significantly differs from the collisionless adiabatic invariant. In a more sophisticated model, we use a random walk representation of Lorentz scattering and calculate the expected increase in energy from the wall movement. This model predicts an inverse relationship between the change in energy and the initial energy. Specifically, it predicts that in the limit of small compression $(\Delta E \ll E_0)$ with many pitch angle scattering collisions $(\tau_{col} \ll T_{comp})$, the change in energy scales with $E_0^{-1/4}$, where $E_0$ is a particle's initial energy, $\tau_{col}$ is the time between collisions, and $T_{comp}$ is the total compression time. We also further confirm the scaling in this limit through a basic computational particle simulation. The inverse relationship allows less energetic particles to experience a greater increase in energy than more energetic ones, resulting in narrower distributions compressed in velocity space. This is the opposite relationship of that described by the well known example of slowly compressing a container of non-interacting particles. This unique relationship and the non-Hamiltonian nature of the system makes this problem interesting to study, particularly because of the possibility of non-thermal features and phase space non-conservation. \\
\indent The paper is organized as follows. In Sec. II, we consider a simple 1-D problem where particles are reflected after traveling a mean free path and identify an invariant. In Sec. III, we describe the implications of our 1-D random walk model and determine how the energy increase scales with a particle's initial energy. Sec. IV features a summary of our study and a discussion of the results.

\section{\label{sec:level1} 1-D Plasma Wall Approximation}
Many Fermi acceleration models can be simply described by interactions of bouncing balls with moving and stationary rigid walls. Our variation is characterized by the inclusion of two species of ions with significantly different masses and the interactions between them. To get an idea of the physics of this system, we start by studying a simpler problem where particles are reflected after traveling a mean free path. This simple plasma wall approximation captures the key physics phenomena.
\subsection{\label{sec:level2} Model and Assumptions}
\indent Consider a 1-D model of an ensemble of ions in a box interacting with a rigid wall moving at constant speed $v_w$, much smaller than any particle speed. A second ensemble of more massive ions is assumed to be nearly stationary in the background and either passes through or sticks to the moving wall as it compresses. The interspecies Coulomb collision time is assumed to be the smallest time scale, followed by the total compression time and then the collision time for the light ions interacting with themselves. This time ordering prevents the light species from thermalizing during compression. In the collisionless limit, particles bounce back and forth between the moving wall and a stationary wall at the other end of the box, separated by distance $L$ and conserving action. In the highly collisional limit, pitch angle scattering is considered to be the primary mechanism of reflecting particles back towards the moving wall. A simple way to model this effect in the 1-D problem is to have particles be reflected after traveling one mean free path, $\lambda_{mfp} = \alpha v^4$, into the box, where $\alpha$ is a constant, and $v$ is the particle speed. The length of the box $L$ is much greater than any particle's mean free path, so most particles are reflected before ever reaching the stationary wall on the other side. Therefore, that wall can be neglected. Fig. 2 shows a graphic of the plasma wall approximation. 
\begin{center}
\includegraphics[scale=0.65]{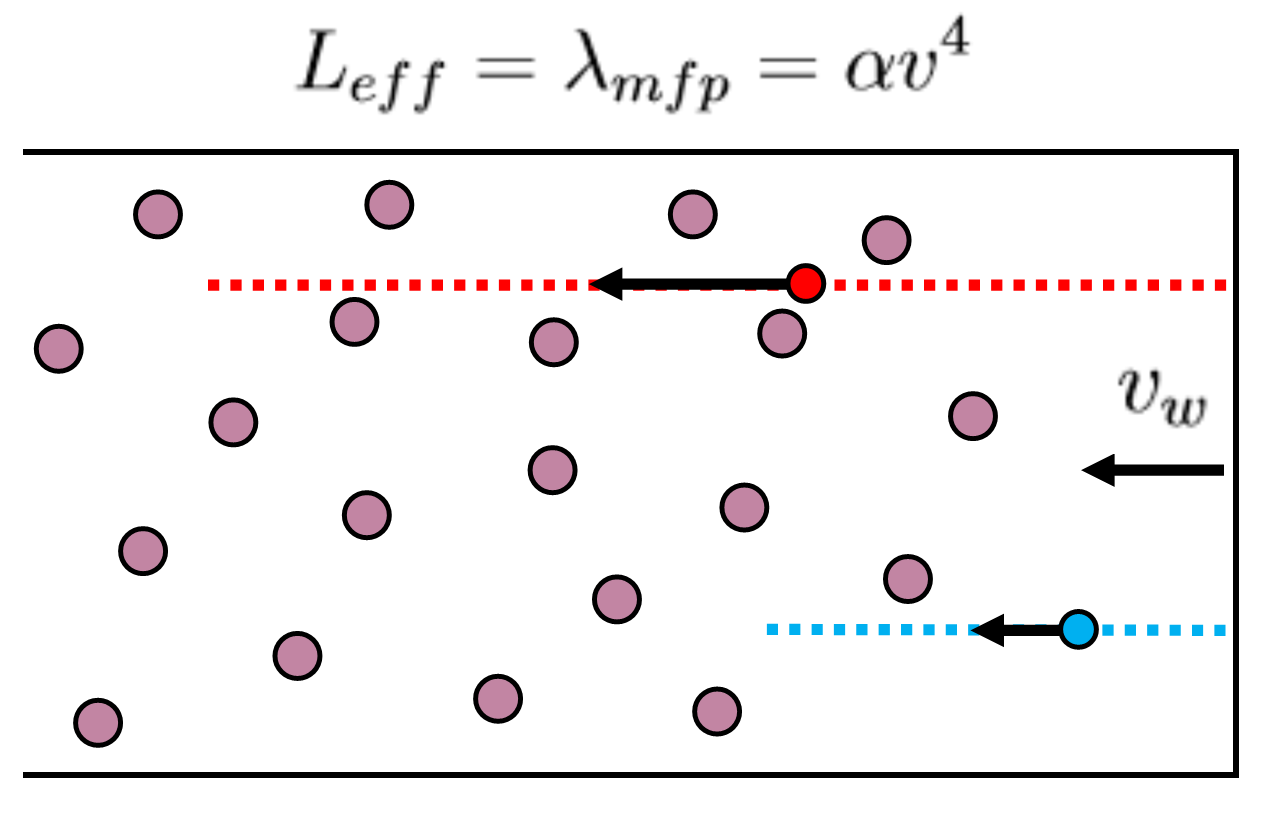} \\
\begin{justify}
Figure 2: Particles interacting with a moving wall while being reflected after traveling a mean free path.
\end{justify}
\end{center}

\subsection{\label{sec:level2} Invariants of Motion}
Key aspects of the pitch angle scattering can be expressed by the 1-D plasma wall model. In particular, the nature of the interaction is non-Hamiltonian, allowing for phase space non-conservation, although the system still exhibits invariants. The adiabatic invariant for collisionless particles being compressed in a 1-D box with width $L$ is the well known 
\begin{eqnarray}
J_1 & = & vL,
\end{eqnarray}
and it turns out that the plasma wall model holds the invariant 
\begin{eqnarray}
J_2 & = & \Delta L + \frac{1}{4}\alpha v^4,
\end{eqnarray}
where $\Delta L$ is the total distance compressed, taken to be negative. The moving wall displacement over a particle bounce time is given by
\begin{eqnarray}
dL & = & -v_w t_{bounce},
\end{eqnarray}
where $t_{bounce} = 2 \alpha v^3$ is the amount of time a particle with speed $v$ spends between interactions with the moving wall. By making this substitution and using the relation $dv = 2v_w$, the equation becomes
\begin{eqnarray}
dL & = & -\alpha v^3 dv.
\end{eqnarray}
Finally, by integrating and rearranging we are left with
\begin{eqnarray}
\Delta L + \frac{1}{4}\alpha v^4 & = & C,
\end{eqnarray}
where $C$ is some constant which we will denote as $J_2$. Although this invariant is a direct result of $v_w$ being a small parameter, it is not formally an adiabatic invariant. Adiabatic invariants are synonymous with action conservation in slowly varying Hamiltonian systems, however this system is non-Hamiltonian, and we will show it does not conserve total phase space. \\
\indent These two invariants have a stark physical difference since particles which conserve $J_1$ will experience a greater acceleration rate if they begin with a greater initial velocity, while the opposite is true for $J_2$. This relationship occurs because in the plasma wall model, fast particles spend a greater amount of time away from the wall since the mean free path scales with $v^4$ which is a phenomena unique to Lorentz scattering. The fact that less energetic particles achieve a greater increase in energy over some compression time causes velocity space compression. \\
\indent To gain some insight on how the particle distribution function evolves as a whole, we can consider that there is some distribution of the amount of time a particle spends between bounces. The evolution of the particle distribution function $f(v,t)$ during some small compression is then described by the advection-diffusion equation
\begin{eqnarray}
\overline{t}\frac{\partial f(v,t)}{\partial t} + 2v_w \frac{\partial f(v,t)}{\partial v} & = & \frac{\delta t^2}{2}\frac{\partial^2 f(v,t)}{\partial t^2},
\end{eqnarray}
where $\overline{t}$ and $\delta t$ are the mean and standard deviation of the bounce time distribution. This equation is derived in a similar way as the Fokker-Planck equation. For both collisional limits, the solution mean follows the respective invariant with variance 
\begin{eqnarray} 
\sigma^2 & = & \frac{\delta t^2 v_w (v - v_0)}{2}. 
\end{eqnarray}
Fig. 3 shows the evolution of a uniform energy distribution during compression as it conserves the invariant $J_2$ with no time variance in the bounce distribution $(\delta t = 0)$. Since the particles with greater energy experience less acceleration, the distribution becomes compressed in velocity space, resulting in a narrower peaked distribution. 
\begin{center}
    \includegraphics[scale=0.8]{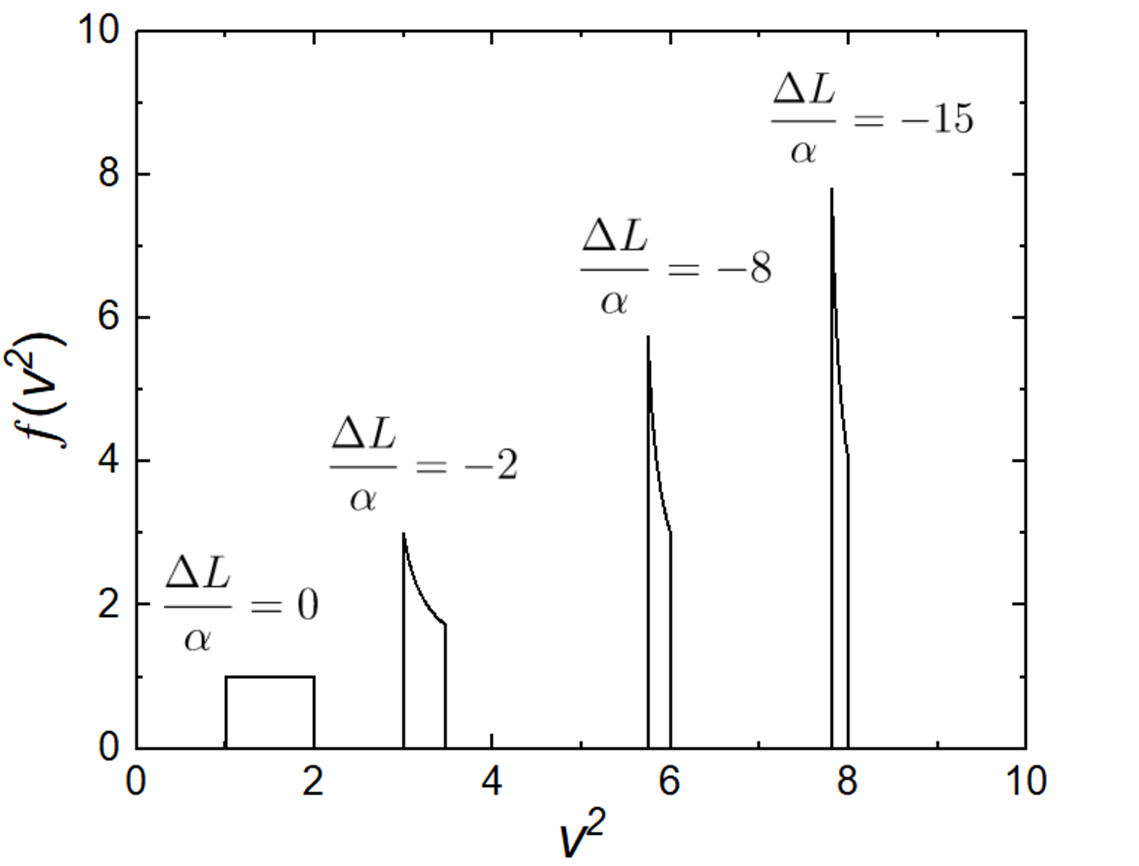} \\
    \begin{justify}
    Figure 3: Evolution of a step energy distribution conserving the invariant $J_2 = \Delta L + \frac{1}{4}\alpha v^4$.
    \end{justify}
\end{center}
\subsection{\label{sec:level2} Phase Space Volume}
Although we have identified compression in velocity space, this does not necessarily mean that total phase space is compressed. As particles gain energy from interacting with the moving wall, they will occupy a greater amount of physical space since the mean free path increases with an increased speed. \\
\indent To understand how these effects compete, consider an initial distribution uniformly distributed in 2-D ($x,v$) phase space between $(v = v_0, v = v_1)$ and $(x = 0, x = \lambda_{mfp})$. When the rigid wall begins to move, particles will gain energy from interacting with this wall. As previously shown, the less energetic particles will experience the greatest increase in speed ($\Delta v_0 > \Delta v_1$). The initial and final phase space volumes are given by
\begin{eqnarray}
P_i & = & \int\limits_{v_0}^{v_1}\alpha v^4 dv 
\end{eqnarray}
and
\begin{eqnarray}
P_f & = & \int\limits_{v_0 + \Delta v_0}^{v_1 + \Delta v_1}\alpha v^4 dv,
\end{eqnarray}
respectively. In the limit of small wall compression ($\Delta v / v << 1$) the change in phase space volume to first order in $\Delta v_{0,1}$ is
\begin{eqnarray}
\Delta P & = & \alpha (v_1^4 \Delta v_1 - v_0^4 \Delta v_0).
\end{eqnarray}
In this limit we can also approximate the form of $\Delta v$ as
\begin{eqnarray}
\Delta v & = & 2v_w \frac{T_{comp}}{2\alpha v^3}.
\end{eqnarray}
Making this substitution into (10) yields
\begin{eqnarray}
\Delta P & = & v_w T_{comp} (v_1 - v_0),
\end{eqnarray}
which clearly is greater than zero. Therefore, the expansion in physical space causes a net phase space volume increase despite the compression in velocity space.
\section{\label{sec:level1} Random Walk Model}
The plasma wall model described in Sec. II of ions reflecting after traveling one mean free path provides insight into the physics of Fermi acceleration with Lorentz scattering, however it fails to fully capture the effect and cannot easily be scaled to three dimensions. A more accurate model of pitch angle scattering can be described with a random walk rather than forcing particle reflection at a mean free path. This is also relatively straightforward to extend from 1-D to higher dimensions. The random walk model turns out to retain the most important feature, which is the inverse relationship between a particle's change in energy and its initial energy. Furthermore, in the limit of many collisions during the compression time, we can estimate the exact form of this relationship.
\subsection{\label{sec:level2} Additional Assumptions}
\indent Particles now take a random walk with the step size equal to one mean free path rather than being reflected. After each particle collision time $\tau_{col} = \alpha v^3$, a particle will either continue on in the same direction or be reflected, exhibiting a random walk. This permits a stochastic component into the system since a particle bounce time is now described by a probability distribution rather than a set time. A graphic of the random walk model is shown in Fig. 4.
\begin{center}
    \includegraphics[scale=0.65]{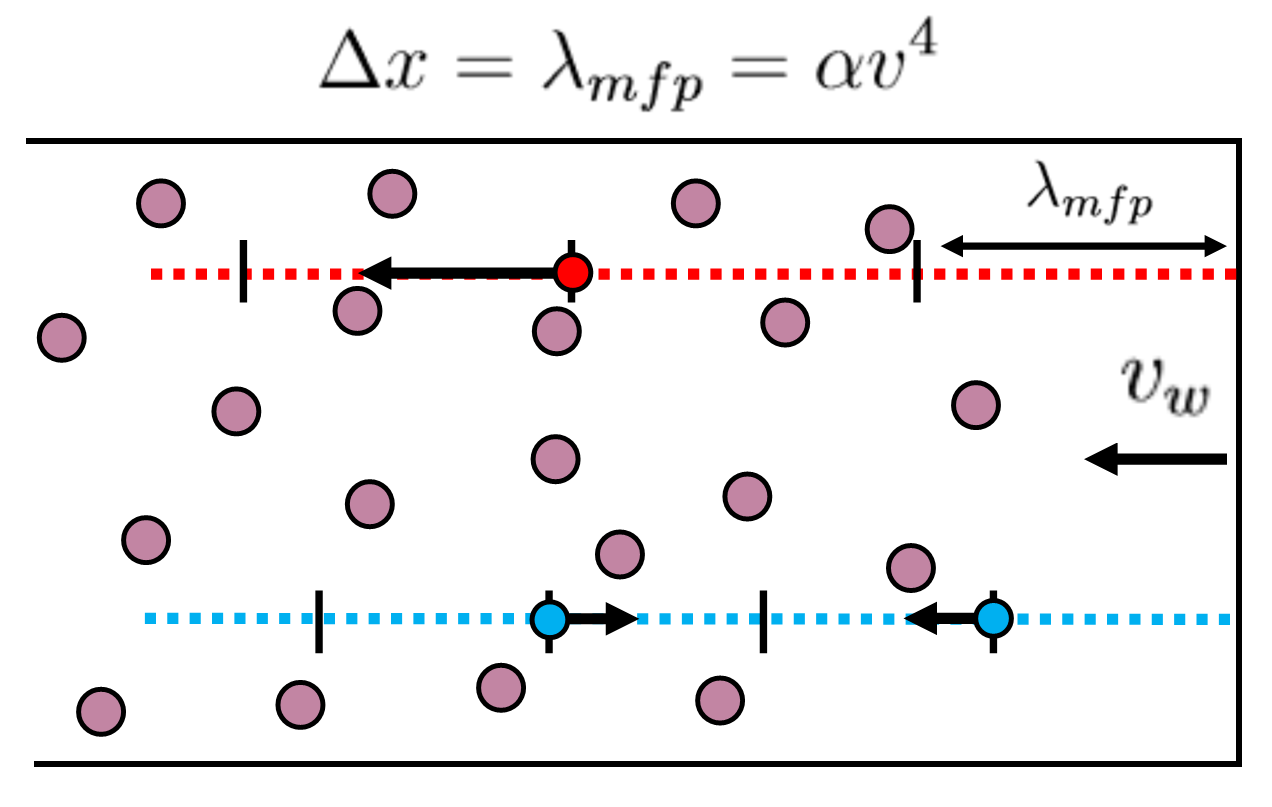} \\
    Figure 4: Particles interacting with a moving wall while taking a 1-D random walk with a step size equal to a collisional mean free path.
\end{center}
\subsection{\label{sec:level2} 1-D Energy Scaling}
\indent In a 1-D random walk of $n$ steps, the expected number of equalizations (or returns to the origin) $r$ in the limit of large $n$ is interestingly given by
\begin{eqnarray}
E[r] & = & \sqrt{\frac{2}{\pi}}\sqrt{n}
\end{eqnarray}
\cite{Feller50, Grinstead97}. The number of steps in our particle system can be expressed in terms of the total compression time $T_{comp}$ and the collision time, $\tau_{col}$ as $n = T_{comp}/\tau_{col}$. Therefore, for small compression $(\Delta v \ll v_0)$ the expected increase in velocity for a particle with initial speed $v_0$ is
\begin{eqnarray}
E[\Delta v] & = & \frac{2\sqrt{2}}{\sqrt{\pi}}\sqrt{\frac{T_{comp}}{\alpha v_0^3}}v_w.
\end{eqnarray}
This yields an unusual dependence of the speed increase with $v^{-3/2}$ or equivalently the energy increase with $E^{-1/4}$. The inverse dependence implies that slower, less energetic particles will receive a greater kick in energy than more energetic particles, resulting in the expected velocity space compression. Fig. 5 shows the evolution of a uniform energy distribution adhering to the expected velocity gain given by (14). Here, we define $T_{comp} = -\Delta L / v_w$.
\begin{center}
    \includegraphics[scale=0.68]{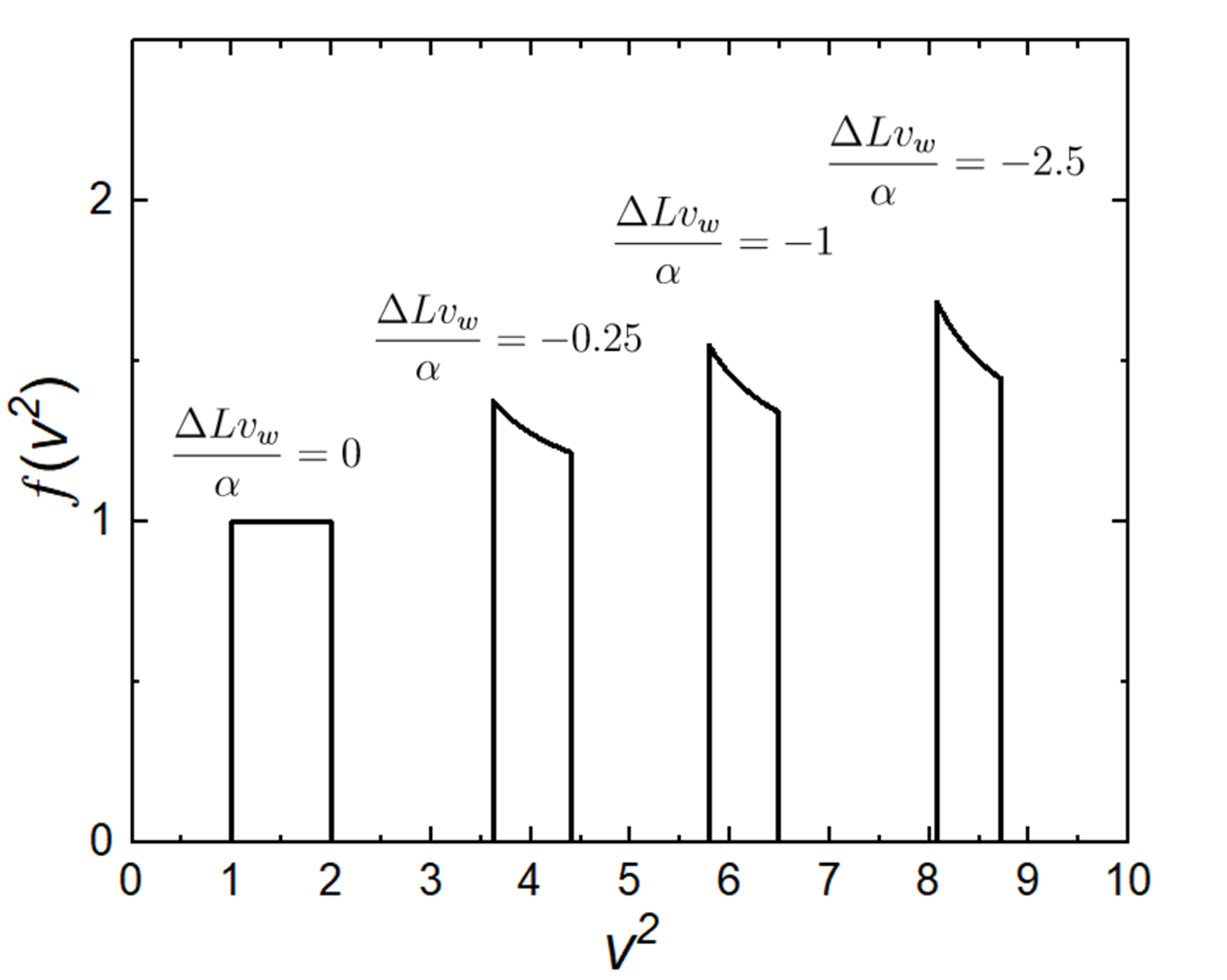} \\
    \begin{justify}
    Figure 5: Evolution of a step energy distribution from compression while following the rate of the expected number of wall interactions given by (14).
    \end{justify}
\end{center}
\indent \indent The compression of the distribution in velocity space is similar to that demonstrated in Fig. 3, although to a lesser degree. This is caused by the random walk model having a weaker inverse scaling between a particle's speed and the acceleration it will experience. In this figure we have also assumed that all particles evolve according to the expected energy gain and have not accounted for the full distribution of energy gains from the random walk model. Nevertheless, we still see the expected compression in velocity space. As the distribution evolves, it becomes narrower and develops a peak in the lower energy regime of the distribution. Since the system is stochastic, the total phase space is consequently also not conserved as in the plasma wall approximation.
\subsection{\label{sec:level2} Higher Dimensions}
\indent For systems of higher dimension, the scaling law given by (13) for the 1-D problem is expected to hold, with the only difference being the constant factor. To demonstrate this, consider an $m$-dimensional spatial system with the wall moving parallel to any one of the dimensions. Suppose that particles randomly walk along a uniform $m$-dimensional lattice with gridpoints separated by a mean free path distance. If each scattering direction is equally likely to occur at a step and a particle takes a total number of $n$ steps, then the total number of these steps expected to be along the direction parallel to the wall's movement is $n/m$. Since the system has translation symmetry in all directions perpendicular to the wall's movement, this setup is equivalent to a 1-D random walk of $n/m$ steps, and therefore the expected number of particle-wall interactions is given by
\begin{eqnarray}
E[r] & = & \sqrt{\frac{2}{\pi m}}\sqrt{n}.
\end{eqnarray}
Understandably, this is less than the strictly 1-D case since particles can now `waste' steps on other degrees of freedom. In a realistic 3-D system, particles are not restricted to a grid so the constant factor in (15) will be different to account for scattering at any spherical angle.
\subsection{\label{sec:level2} Numerical Results}
A simple particle simulation was written to investigate the effect of Coulomb pitch angle scattering in the 3-D random walk case. A total of $10^5$ particles were initialized at the surface of a moving, rigid wall with velocities directed away from the wall. The particles were divided into ten groups with different initial velocities in order to determine how the speed increase scaled with the initial speed. There was no stationary wall implemented on the other side of the simulation domain, so pitch angle scattering was the only mechanism responsible for turning particles back towards the moving wall. Collisions were simulated by randomly changing a particle's pitch angle every time it traveled one mean free path, $\lambda_{mfp} = \alpha v^4$. Fig. 6 shows the average change in velocity for each subgroup as a function of their initial velocity after a compression time of about $5000$ collisions for the least energetic group of particles.
\begin{center}
    \includegraphics[scale=0.57]{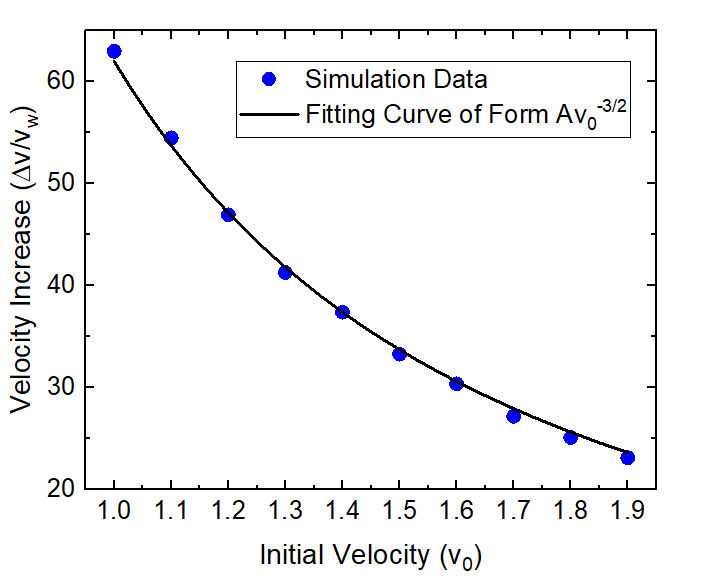} \\
    \begin{justify}
    Figure 6: Velocity increase following compression as a function of initial velocity. In this simulation, $\alpha = 0.1$, $v_w = 10^{-9}v_0$, and $T_{comp} = 5000 \tau_{col_{v_0}}$.
    \end{justify}
\end{center}
\indent \indent Clearly the result shows an inverse relationship between the two variables as shown earlier in the 1-D case. The scaling is near $\Delta v \sim v_0^{-3/2}$ (or equivalently $\Delta E \sim E_0^{-1/4}$) as shown by the fitting curve, which is also consistent with our analytic prediction from earlier in the section. This inverse scaling is the key to obtaining the non-thermal peaked distributions shown in Fig. 3 and Fig. 5 from velocity space compression. 
\\
\\
\section{\label{sec:level1} Summary and Discussion}
We presented a simple model for a two species ion ensemble interacting with a moving wall while also undergoing pitch angle scattering. We predicted an interesting inverse relationship between the change in energy from compression and the initial particle energy. It follows that less energetic particles experience greater acceleration, resulting in compression of the particle distribution in velocity space. This velocity space compression could generate potentially favorable peaked energy distributions as shown in Fig. 3 and Fig. 5. \\
\indent The non-thermal phenomena predicted by these models could be of interest to various areas of plasma physics since we are considering Lorentz scattering of charged particles. In particular, due to the mass difference between the two species, the model could describe some aspects of p$^{11}$B interactions. This is a potential fuel source for aneutronic fusion \cite{DAWSON1981453} and the velocity space compression could provide a mechanism for obtaining favorable proton energy distributions to increase fusion reactivity \cite{Becker87, Nevins_2000, Sikora16}. The moving wall in this case could represent either a physical wall compressing, or a moving magnetic field structure in a magnetic mirror confinement setup \cite{Nevins98}. The model could also be used to describe a variation of cosmic ray acceleration, where the Lorentz scattering mean free path is small compared to the system size. Identifying acceleration mechanisms in astrophysical settings using Fermi acceleration models is an ongoing area of study \cite{Malkov2001, Fox17, Veltri04} and the phenomena outlined in this paper may be applicable to the field. \\
\indent We opted for simplicity in our models to isolate the important effects described in this paper. However, in real plasmas there are more complex features such as collisions within a species itself which allows ions to thermalize. Thermalization could significantly dampen any velocity space compression, which is why we examined the limit where interspecies collisions dominate over the light ion collisions with themselves. In this limit we have identified some unique effects of Fermi acceleration with Lorentz scattering. The most interesting aspect of the results is the potential for non-thermal, non-Hamiltonian features in the compression due to the non-conservation of phase space. \\
\begin{acknowledgments}
The authors are thankful to Elijah Kolmes, Ian Ochs, and Tal Rubin for helpful conversations. This work is supported by NSF PHY-1805316 and NNSA DE-SC0021248.
\end{acknowledgments}

\nocite{*}

\bibliography{apssamp}

\providecommand{\noopsort}[1]{}\providecommand{\singleletter}[1]{#1}%
\begin{thebibliography}{25}%
\makeatletter
\providecommand \@ifxundefined [1]{%
 \@ifx{#1\undefined}
}%
\providecommand \@ifnum [1]{%
 \ifnum #1\expandafter \@firstoftwo
 \else \expandafter \@secondoftwo
 \fi
}%
\providecommand \@ifx [1]{%
 \ifx #1\expandafter \@firstoftwo
 \else \expandafter \@secondoftwo
 \fi
}%
\providecommand \natexlab [1]{#1}%
\providecommand \enquote  [1]{``#1''}%
\providecommand \bibnamefont  [1]{#1}%
\providecommand \bibfnamefont [1]{#1}%
\providecommand \citenamefont [1]{#1}%
\providecommand \href@noop [0]{\@secondoftwo}%
\providecommand \href [0]{\begingroup \@sanitize@url \@href}%
\providecommand \@href[1]{\@@startlink{#1}\@@href}%
\providecommand \@@href[1]{\endgroup#1\@@endlink}%
\providecommand \@sanitize@url [0]{\catcode `\\12\catcode `\$12\catcode
  `\&12\catcode `\#12\catcode `\^12\catcode `\_12\catcode `\%12\relax}%
\providecommand \@@startlink[1]{}%
\providecommand \@@endlink[0]{}%
\providecommand \url  [0]{\begingroup\@sanitize@url \@url }%
\providecommand \@url [1]{\endgroup\@href {#1}{\urlprefix }}%
\providecommand \urlprefix  [0]{URL }%
\providecommand \Eprint [0]{\href }%
\providecommand \doibase [0]{https://doi.org/}%
\providecommand \selectlanguage [0]{\@gobble}%
\providecommand \bibinfo  [0]{\@secondoftwo}%
\providecommand \bibfield  [0]{\@secondoftwo}%
\providecommand \translation [1]{[#1]}%
\providecommand \BibitemOpen [0]{}%
\providecommand \bibitemStop [0]{}%
\providecommand \bibitemNoStop [0]{.\EOS\space}%
\providecommand \EOS [0]{\spacefactor3000\relax}%
\providecommand \BibitemShut  [1]{\csname bibitem#1\endcsname}%
\let\auto@bib@innerbib\@empty
\bibitem [{\citenamefont {Fermi}(1949)}]{Fermi49}%
  \BibitemOpen
  \bibfield  {author} {\bibinfo {author} {\bibfnamefont {E.}~\bibnamefont
  {Fermi}},\ }\href@noop {} {\bibfield  {journal} {\bibinfo  {journal} {Phys.\
  Rev.}\ }\textbf {\bibinfo {volume} {75}},\ \bibinfo {pages} {1169} (\bibinfo
  {year} {1949})}\BibitemShut {NoStop}%
\bibitem [{\citenamefont {Ulam}(1961)}]{Ulam61}%
  \BibitemOpen
  \bibfield  {author} {\bibinfo {author} {\bibfnamefont {S.}~\bibnamefont
  {Ulam}},\ }\bibfield  {title} {\bibinfo {title} {Proceedings of the 4$^{th}$
  berkeley symposium on mathematical statistics and probability}\ }(\bibinfo
  {publisher} {California University Press},\ \bibinfo {address} {Berkeley,
  CA},\ \bibinfo {year} {1961})\ p.\ \bibinfo {pages} {315}\BibitemShut
  {NoStop}%
\bibitem [{\citenamefont {Gelfriech}\ \emph {et~al.}(2012)\citenamefont
  {Gelfriech}, \citenamefont {Rom-Kedar},\ and\ \citenamefont
  {Turaev}}]{Gelfriech12}%
  \BibitemOpen
  \bibfield  {author} {\bibinfo {author} {\bibfnamefont {V.}~\bibnamefont
  {Gelfriech}}, \bibinfo {author} {\bibfnamefont {V.}~\bibnamefont
  {Rom-Kedar}},\ and\ \bibinfo {author} {\bibfnamefont {D.}~\bibnamefont
  {Turaev}},\ }\href@noop {} {\bibfield  {journal} {\bibinfo  {journal}
  {Chaos}\ }\textbf {\bibinfo {volume} {22}},\ \bibinfo {pages} {033116}
  (\bibinfo {year} {2012})}\BibitemShut {NoStop}%
\bibitem [{\citenamefont {Zhou}(2020)}]{Zhou2020}%
  \BibitemOpen
  \bibfield  {author} {\bibinfo {author} {\bibfnamefont {J.}~\bibnamefont
  {Zhou}},\ }\href {https://doi.org/10.1088/1361-6544/ab60d7} {\bibfield
  {journal} {\bibinfo  {journal} {Nonlinearity}\ }\textbf {\bibinfo {volume}
  {33}},\ \bibinfo {pages} {1542} (\bibinfo {year} {2020})}\BibitemShut
  {NoStop}%
\bibitem [{\citenamefont {Lenz}\ \emph {et~al.}(2008)\citenamefont {Lenz},
  \citenamefont {Diakonos},\ and\ \citenamefont {Schmelcher}}]{Lenz08}%
  \BibitemOpen
  \bibfield  {author} {\bibinfo {author} {\bibfnamefont {F.}~\bibnamefont
  {Lenz}}, \bibinfo {author} {\bibfnamefont {F.~K.}\ \bibnamefont {Diakonos}},\
  and\ \bibinfo {author} {\bibfnamefont {P.}~\bibnamefont {Schmelcher}},\
  }\href {https://doi.org/10.1103/PhysRevLett.100.014103} {\bibfield  {journal}
  {\bibinfo  {journal} {Phys. Rev. Lett.}\ }\textbf {\bibinfo {volume} {100}},\
  \bibinfo {pages} {014103} (\bibinfo {year} {2008})}\BibitemShut {NoStop}%
\bibitem [{\citenamefont {Leonel}\ \emph {et~al.}(2009)\citenamefont {Leonel},
  \citenamefont {Oliveira},\ and\ \citenamefont {Loskutov}}]{Leonel09}%
  \BibitemOpen
  \bibfield  {author} {\bibinfo {author} {\bibfnamefont {E.~D.}\ \bibnamefont
  {Leonel}}, \bibinfo {author} {\bibfnamefont {D.~F.~M.}\ \bibnamefont
  {Oliveira}},\ and\ \bibinfo {author} {\bibfnamefont {A.}~\bibnamefont
  {Loskutov}},\ }\href {https://doi.org/10.1063/1.3227740} {\bibfield
  {journal} {\bibinfo  {journal} {Chaos}\ }\textbf {\bibinfo {volume} {19}},\
  \bibinfo {pages} {033142} (\bibinfo {year} {2009})}\BibitemShut {NoStop}%
\bibitem [{\citenamefont {Jarzynski}(1993)}]{Jarzynski93}%
  \BibitemOpen
  \bibfield  {author} {\bibinfo {author} {\bibfnamefont {C.}~\bibnamefont
  {Jarzynski}},\ }\href@noop {} {\bibfield  {journal} {\bibinfo  {journal}
  {Phys.\ Rev. E}\ }\textbf {\bibinfo {volume} {48}},\ \bibinfo {pages} {6}
  (\bibinfo {year} {1993})}\BibitemShut {NoStop}%
\bibitem [{\citenamefont {Loskutov}\ \emph {et~al.}(2000)\citenamefont
  {Loskutov}, \citenamefont {Ryabov},\ and\ \citenamefont
  {Akinshin}}]{Loskutov2000}%
  \BibitemOpen
  \bibfield  {author} {\bibinfo {author} {\bibfnamefont {A.}~\bibnamefont
  {Loskutov}}, \bibinfo {author} {\bibfnamefont {A.~B.}\ \bibnamefont
  {Ryabov}},\ and\ \bibinfo {author} {\bibfnamefont {L.~G.}\ \bibnamefont
  {Akinshin}},\ }\href {https://doi.org/10.1088/0305-4470/33/44/309} {\bibfield
   {journal} {\bibinfo  {journal} {J. Phys. A: Math. Gen.}\ }\textbf {\bibinfo
  {volume} {33}},\ \bibinfo {pages} {7973} (\bibinfo {year}
  {2000})}\BibitemShut {NoStop}%
\bibitem [{\citenamefont {Gelfreich}\ and\ \citenamefont
  {Turaev}(2008)}]{Gelfreich2008}%
  \BibitemOpen
  \bibfield  {author} {\bibinfo {author} {\bibfnamefont {V.}~\bibnamefont
  {Gelfreich}}\ and\ \bibinfo {author} {\bibfnamefont {D.}~\bibnamefont
  {Turaev}},\ }\href {https://doi.org/10.1088/1751-8113/41/21/212003}
  {\bibfield  {journal} {\bibinfo  {journal} {J. Phys. A: Math. Theor.}\
  }\textbf {\bibinfo {volume} {41}},\ \bibinfo {pages} {212003} (\bibinfo
  {year} {2008})}\BibitemShut {NoStop}%
\bibitem [{\citenamefont {Leonel}\ and\ \citenamefont
  {McClintock}(2005)}]{Leonel_2005}%
  \BibitemOpen
  \bibfield  {author} {\bibinfo {author} {\bibfnamefont {E.~D.}\ \bibnamefont
  {Leonel}}\ and\ \bibinfo {author} {\bibfnamefont {P.~V.~E.}\ \bibnamefont
  {McClintock}},\ }\href {https://doi.org/10.1088/0305-4470/38/4/004}
  {\bibfield  {journal} {\bibinfo  {journal} {J. Phys. A: Math. Gen.}\ }\textbf
  {\bibinfo {volume} {38}},\ \bibinfo {pages} {823} (\bibinfo {year}
  {2005})}\BibitemShut {NoStop}%
\bibitem [{\citenamefont {Silveira}\ \emph {et~al.}(2021)\citenamefont
  {Silveira}, \citenamefont {Alves}, \citenamefont {Leonel},\ and\
  \citenamefont {Ladeira}}]{Silveira21}%
  \BibitemOpen
  \bibfield  {author} {\bibinfo {author} {\bibfnamefont {F.~A.~O.}\
  \bibnamefont {Silveira}}, \bibinfo {author} {\bibfnamefont {S.~G.}\
  \bibnamefont {Alves}}, \bibinfo {author} {\bibfnamefont {E.~D.}\ \bibnamefont
  {Leonel}},\ and\ \bibinfo {author} {\bibfnamefont {D.~G.}\ \bibnamefont
  {Ladeira}},\ }\href {https://doi.org/10.1103/PhysRevE.103.062205} {\bibfield
  {journal} {\bibinfo  {journal} {Phys. Rev. E}\ }\textbf {\bibinfo {volume}
  {103}},\ \bibinfo {pages} {062205} (\bibinfo {year} {2021})}\BibitemShut
  {NoStop}%
\bibitem [{\citenamefont {Chandran}(2000)}]{Chandran00}%
  \BibitemOpen
  \bibfield  {author} {\bibinfo {author} {\bibfnamefont {B.~D.~G.}\
  \bibnamefont {Chandran}},\ }\href@noop {} {\bibfield  {journal} {\bibinfo
  {journal} {Phys. Rev. Lett.}\ }\textbf {\bibinfo {volume} {85}},\ \bibinfo
  {pages} {4656} (\bibinfo {year} {2000})}\BibitemShut {NoStop}%
\bibitem [{\citenamefont {Selkowitz}\ and\ \citenamefont
  {Blackman}(2004)}]{Selkowitz04}%
  \BibitemOpen
  \bibfield  {author} {\bibinfo {author} {\bibfnamefont {R.}~\bibnamefont
  {Selkowitz}}\ and\ \bibinfo {author} {\bibfnamefont {E.~G.}\ \bibnamefont
  {Blackman}},\ }\href@noop {} {\bibfield  {journal} {\bibinfo  {journal}
  {MNRAS}\ }\textbf {\bibinfo {volume} {354}},\ \bibinfo {pages} {870}
  (\bibinfo {year} {2004})}\BibitemShut {NoStop}%
\bibitem [{\citenamefont {Liu}\ \emph {et~al.}(2017)\citenamefont {Liu},
  \citenamefont {Lu}, \citenamefont {Angelopoulos}, \citenamefont {Hietala},\
  and\ \citenamefont {Wilson~III}}]{Liu17}%
  \BibitemOpen
  \bibfield  {author} {\bibinfo {author} {\bibfnamefont {T.~Z.}\ \bibnamefont
  {Liu}}, \bibinfo {author} {\bibfnamefont {S.}~\bibnamefont {Lu}}, \bibinfo
  {author} {\bibfnamefont {V.}~\bibnamefont {Angelopoulos}}, \bibinfo {author}
  {\bibfnamefont {H.}~\bibnamefont {Hietala}},\ and\ \bibinfo {author}
  {\bibfnamefont {L.~B.}\ \bibnamefont {Wilson~III}},\ }\href@noop {}
  {\bibfield  {journal} {\bibinfo  {journal} {J. Geophys. Res. Space Phys.}\
  }\textbf {\bibinfo {volume} {122}},\ \bibinfo {pages} {9248} (\bibinfo {year}
  {2017})}\BibitemShut {NoStop}%
\bibitem [{\citenamefont {Scott}\ \emph {et~al.}(1978)\citenamefont {Scott},
  \citenamefont {Cocke}, \citenamefont {Chevalier},\ and\ \citenamefont
  {Wentzel}}]{Scott78}%
  \BibitemOpen
  \bibfield  {author} {\bibinfo {author} {\bibfnamefont {J.}~\bibnamefont
  {Scott}}, \bibinfo {author} {\bibfnamefont {W.}~\bibnamefont {Cocke}},
  \bibinfo {author} {\bibfnamefont {R.}~\bibnamefont {Chevalier}},\ and\
  \bibinfo {author} {\bibfnamefont {D.}~\bibnamefont {Wentzel}},\ }\href@noop
  {} {\bibfield  {journal} {\bibinfo  {journal} {Astrophys. Space Sci.}\
  }\textbf {\bibinfo {volume} {53}},\ \bibinfo {pages} {421} (\bibinfo {year}
  {1978})}\BibitemShut {NoStop}%
\bibitem [{\citenamefont {Feller}(1950)}]{Feller50}%
  \BibitemOpen
  \bibfield  {author} {\bibinfo {author} {\bibfnamefont {W.}~\bibnamefont
  {Feller}},\ }\href@noop {} {\emph {\bibinfo {title} {An Introduction to
  Probability Theory and Its Applications}}},\ Vol.~\bibinfo {volume} {1}\
  (\bibinfo  {publisher} {Wiley},\ \bibinfo {address} {New York},\ \bibinfo
  {year} {1950})\BibitemShut {NoStop}%
\bibitem [{\citenamefont {Grinstead}\ and\ \citenamefont
  {Snell}(1997)}]{Grinstead97}%
  \BibitemOpen
  \bibfield  {author} {\bibinfo {author} {\bibfnamefont {C.}~\bibnamefont
  {Grinstead}}\ and\ \bibinfo {author} {\bibfnamefont {J.~L.}\ \bibnamefont
  {Snell}},\ }\href@noop {} {\emph {\bibinfo {title} {Introduction to
  Probability}}}\ (\bibinfo  {publisher} {American Mathematical Society},\
  \bibinfo {address} {Providence, RI},\ \bibinfo {year} {1997})\BibitemShut
  {NoStop}%
\bibitem [{\citenamefont {Dawson}(1981)}]{DAWSON1981453}%
  \BibitemOpen
  \bibfield  {author} {\bibinfo {author} {\bibfnamefont {J.}~\bibnamefont
  {Dawson}},\ }\bibfield  {title} {\bibinfo {title} {Advanced fusion
  reactors},\ }in\ \href
  {https://doi.org/https://doi.org/10.1016/B978-0-12-685241-7.50013-X} {\emph
  {\bibinfo {booktitle} {Fusion}}},\ Vol.~\bibinfo {volume} {1},\ \bibinfo
  {editor} {edited by\ \bibinfo {editor} {\bibfnamefont {E.}~\bibnamefont
  {Teller}}}\ (\bibinfo  {publisher} {Academic Press},\ \bibinfo {address} {New
  York},\ \bibinfo {year} {1981})\ pp.\ \bibinfo {pages} {453--501}\BibitemShut
  {NoStop}%
\bibitem [{\citenamefont {Becker}\ \emph {et~al.}(1987)\citenamefont {Becker},
  \citenamefont {Rolfs},\ and\ \citenamefont {Trautvetter}}]{Becker87}%
  \BibitemOpen
  \bibfield  {author} {\bibinfo {author} {\bibfnamefont {H.~W.}\ \bibnamefont
  {Becker}}, \bibinfo {author} {\bibfnamefont {C.}~\bibnamefont {Rolfs}},\ and\
  \bibinfo {author} {\bibfnamefont {H.~P.}\ \bibnamefont {Trautvetter}},\
  }\href@noop {} {\bibfield  {journal} {\bibinfo  {journal} {Z. Phys. A}\
  }\textbf {\bibinfo {volume} {327}},\ \bibinfo {pages} {341} (\bibinfo {year}
  {1987})}\BibitemShut {NoStop}%
\bibitem [{\citenamefont {Nevins}\ and\ \citenamefont
  {Swain}(2000)}]{Nevins_2000}%
  \BibitemOpen
  \bibfield  {author} {\bibinfo {author} {\bibfnamefont {W.~M.}\ \bibnamefont
  {Nevins}}\ and\ \bibinfo {author} {\bibfnamefont {R.}~\bibnamefont {Swain}},\
  }\href {https://doi.org/10.1088/0029-5515/40/4/310} {\bibfield  {journal}
  {\bibinfo  {journal} {Nucl. Fusion}\ }\textbf {\bibinfo {volume} {40}},\
  \bibinfo {pages} {865} (\bibinfo {year} {2000})}\BibitemShut {NoStop}%
\bibitem [{\citenamefont {Sikora}\ and\ \citenamefont
  {Weller}(2016)}]{Sikora16}%
  \BibitemOpen
  \bibfield  {author} {\bibinfo {author} {\bibfnamefont {M.~H.}\ \bibnamefont
  {Sikora}}\ and\ \bibinfo {author} {\bibfnamefont {H.~R.}\ \bibnamefont
  {Weller}},\ }\href@noop {} {\bibfield  {journal} {\bibinfo  {journal}
  {Journal of Fusion Energy}\ }\textbf {\bibinfo {volume} {35}},\ \bibinfo
  {pages} {538} (\bibinfo {year} {2016})}\BibitemShut {NoStop}%
\bibitem [{\citenamefont {Nevins}(1998)}]{Nevins98}%
  \BibitemOpen
  \bibfield  {author} {\bibinfo {author} {\bibfnamefont {W.~M.}\ \bibnamefont
  {Nevins}},\ }\href@noop {} {\bibfield  {journal} {\bibinfo  {journal} {J.
  Fusion Energ.}\ }\textbf {\bibinfo {volume} {17}},\ \bibinfo {pages} {25}
  (\bibinfo {year} {1998})}\BibitemShut {NoStop}%
\bibitem [{\citenamefont {Malkov}\ and\ \citenamefont
  {Drury}(2001)}]{Malkov2001}%
  \BibitemOpen
  \bibfield  {author} {\bibinfo {author} {\bibfnamefont {M.~A.}\ \bibnamefont
  {Malkov}}\ and\ \bibinfo {author} {\bibfnamefont {L.~O.}\ \bibnamefont
  {Drury}},\ }\href {https://doi.org/10.1088/0034-4885/64/4/201} {\bibfield
  {journal} {\bibinfo  {journal} {Rep. Prog. Phys.}\ }\textbf {\bibinfo
  {volume} {64}},\ \bibinfo {pages} {429} (\bibinfo {year} {2001})}\BibitemShut
  {NoStop}%
\bibitem [{\citenamefont {Fox}\ \emph {et~al.}(2017)\citenamefont {Fox},
  \citenamefont {Park}, \citenamefont {Deng}, \citenamefont {Fiksel},
  \citenamefont {Spitkovsky},\ and\ \citenamefont {Bhattacharjee}}]{Fox17}%
  \BibitemOpen
  \bibfield  {author} {\bibinfo {author} {\bibfnamefont {W.}~\bibnamefont
  {Fox}}, \bibinfo {author} {\bibfnamefont {J.}~\bibnamefont {Park}}, \bibinfo
  {author} {\bibfnamefont {W.}~\bibnamefont {Deng}}, \bibinfo {author}
  {\bibfnamefont {G.}~\bibnamefont {Fiksel}}, \bibinfo {author} {\bibfnamefont
  {A.}~\bibnamefont {Spitkovsky}},\ and\ \bibinfo {author} {\bibfnamefont
  {A.}~\bibnamefont {Bhattacharjee}},\ }\href
  {https://doi.org/10.1063/1.4993204} {\bibfield  {journal} {\bibinfo
  {journal} {Phys. Plasmas}\ }\textbf {\bibinfo {volume} {24}},\ \bibinfo
  {pages} {092901} (\bibinfo {year} {2017})}\BibitemShut {NoStop}%
\bibitem [{\citenamefont {Veltri}\ and\ \citenamefont
  {Carbone}(2004)}]{Veltri04}%
  \BibitemOpen
  \bibfield  {author} {\bibinfo {author} {\bibfnamefont {A.}~\bibnamefont
  {Veltri}}\ and\ \bibinfo {author} {\bibfnamefont {V.}~\bibnamefont
  {Carbone}},\ }\href {https://doi.org/10.1103/PhysRevLett.92.143901}
  {\bibfield  {journal} {\bibinfo  {journal} {Phys. Rev. Lett.}\ }\textbf
  {\bibinfo {volume} {92}},\ \bibinfo {pages} {143901} (\bibinfo {year}
  {2004})}\BibitemShut {NoStop}%
\end{thebibliography}%

\end{document}